# Depletion of fossil fuel reserves and projections of CO2 concentration in the Earth atmosphere

*Daniele Mazza and Enrico Canuto*
*Politecnico di Torino, Former Faculty, Turin, Italy*
*mazzad50@gmail.com*
*enrico.canuto@formerfaculty.polito.it*

**Abstract**

The paper has been suggested by two observations: 1) the atmospheric $CO_2$ growth rate is smaller than that ascribed to the emission of fossil fuels combustion, 2) the fossil fuel reserves are finite.

The first observation has lead the way to a simple kinetic mode, based on the balance of 1) land/ocean $CO_2$ absorption and 2) $CO_2$ anthropogenic emission limited solely by depletion of the present day fossil-fuel reserves, in a business-as-usual scenario.

The second observation has suggested to extrapolate past $CO_2$ emissions by fossil fuel combustion in the future years up to 2200 CE, by constraining emissions to the physical limits of reserves availability. The Meixner curve (hyperbolic secant distribution) has been used to model the pathway of resource exploitation for the three main classes of fossil fuels, crude oil, natural gas and coal.

The kinetic model, driven by the extrapolated emissions, has been employed to project the $CO_2$ atmospheric concentration due to fossil fuel combustion close to the zero-reserve epoch. The result is just the output of simple models tuned on well-known experimental data. Error analysis of literature data provides the method robustness and the relevant uncertainty band.

Contribution of other greenhouse gases like methane and nitrous oxide has been neglected, since their emissions cannot be projected with the paper methodology (they do not derive from fossil reserves). Notwithstanding this limitation, paper results clearly demonstrate that some of the IPCC projections of the $CO_2$ concentration are largely overestimated if compared to the physical limits of fossil fuel exploitation.

**Acronyms**
CAT Climate Action Tracker
CD carbon dioxide ($CO_2$)
GCB Global Carbon Budget
GHG Greenhouse gas
$GtCO_2$ Gigatonnes of carbon dioxide
IPCC Intergovernmental Panel on Climate Change
NA Not applicable
OWID Our World In Data

ppm part per million
RMS Root mean square
SSP Shared socioeconomic pathway

# 1  Introduction

Future greenhouse gas (GHG) emissions are the subject of extensive research, in relationships with the demanding international pledges about net zero 2050 emissions [20]. The Intergovernmental Panel on Climate Change (IPCC) already in 1990 developed long-term emission scenarios up to 2100 [21]. These different scenarios have been widely used in the analysis of possible climate change, its impact and options to mitigate the change.

In 2000, IPCC issued a Special report on Emissions Scenarios, trying to focus on complex driving forces such as demographic development, socioeconomic development, and technological change [22]. Recently the Climate Action Tracker [29] has summarized in a chart the projections of different policies aiming at reducing, or at least slowing down, the atmospheric concentration growth (in part per million) of $CO_2$ (sometimes denoted by CD, from carbon dioxide) and other greenhouse gases, in an effort of mitigating the relevant effects on future global warming. Revised scenarios and their projections are included in the IPCC Sixth Assessment Report [34] and will be briefly analyzed in Section 4.2.

Surprisingly, all the above cited scenarios and simulation packages (see [31], [32], [38], [40] and the Annex II of [34]) do not include a limiting factor for the fossil fuel emissions, like their intrinsic reserve. The term 'reserve' referred to fossil fuels is marginally cited three times in two figures of the Chapter 5 of [34], dealing with the global carbon cycle.

Present day growths in population and energy per capita are mainly based on fossil fuels as energy sources. Fossil fuels are hydrocarbon-containing materials, which formed naturally on the Earth's surface from the remains of dead plants and animals. The main three categories of fossil fuels are coal, crude oil, natural gas. Fossil fuels may be burned for providing heat for direct use (say for cooking or heating), for powering engines (such as internal combustion engines in motor vehicles), or for generating electricity. The principal origin of fossil fuels is the anaerobic decomposition of buried dead organisms, containing organic molecules synthesized by ancient photosynthesis. Transition from these source materials to high-carbon fossil fuels typically requires geological processes of millions of years, like in the carboniferous geological period, which spanned from 360 to 300 millions years ago. For this reason fossil fuels are typically non-renewable resources.

The main goal of the paper is to predict the future atmospheric $CO_2$ concentration (the annual mean) on the basis of historical data and fossil-fuel reserves (2020 estimates). The paper considers the $CO_2$ concentration as measured by Mauna Loa observatory from 1958 (the so-called Keeling curve [30]), thus neglecting other atmospheric GHG like methane as they depend on different emission sources to and removal mechanisms from atmosphere.

$CO_2$ is dynamically exchanged among atmosphere, biomass, soil and ocean [1], [23]. A big deal of $CO_2$ is taken out from atmosphere by photosynthesis (around 120 $GtCO_2$/year), but at the same time half of this goes back to air by plant emission during night and daily animal breathing. Part of the remainder goes to soil after plant death and, due to bacterial fermentation, again into the air, and part is washed away as $CO_3^{2-}$ by surface fresh water into sea. An estimate of the net exchange can be performed by computerized models. The results of such different models have been recently collected and averaged by the Global Carbon Budget project (see [4] and [12]).

Before industrial era (conventionally starting at 1750 CE), ice core proxy data of the last 2000 years show the atmospheric $CO_2$ concentration (in annual mean) fluctuating around a mean value of about 280 ppm, then since the early XIX century slowly increasing in the average up to World War II and then steadily increasing since the 50s of the past century when systematic and reliable direct measurements were provided by the Mauna Loa observatory. As a result, the periodic annual exchange between soil, ocean and atmosphere is accompanied by an atmospheric increase, small if compared with the volume of the annual exchange, which progressively accumulates. The main contribution to such an inflow has been allotted to $CO_2$ emission of the fossil fuel combustion.

Actually things look not so simple, since the annual estimated $CO_2$ emission flow [ppm/y] happens to be larger than the corresponding atmospheric concentration increase, implying that part of the accumulated $CO_2$ is re-absorbed by soil and ocean as they behave like huge carbon sinks. But how long does it take a perturbation of the atmospheric $CO_2$ concentration to be re-absorbed by soil and ocean sinks? This is a fundamental question to be answered in Section 2 in order to predict the future concentration on the basis of the predicted fossil fuel emissions. The relevant absorption time period (known as *time constant* in the dynamic system field [11] and *relaxation time* in physics and chemistry [10]) should not be confused, as pointed out in [23] and [24], with the CO2 *residence time* of the annual land/ocean-atmosphere two-way, zero-mean flow. In terms of dynamic systems, the latter is nothing else than the $CO_2$ atmospheric *transport delay,* like that of a fluid along a pipe. Actually, both time intervals are referred by Archer [23], pages 112-113, as *lifetimes:* the residence time as the lifetime of the *annual exchange flux* and time constant as the lifetime of the *net flux*. The origin of both time intervals is clearly different: chemistry kinetics at the boundary between land/ocean and atmosphere in the former case, atmospheric transport mechanisms in the latter case.

A clearer equivalence between *lifetime* and the $CO_2$ *time constant* $\tau$, to be estimated in this paper, is provided by [33], page 2-3, as the definition postulates the 1st order differential equation

$$\dot{x}(t) = -\tau^{-1} x(t) + u(t), \quad x(t_0) = x_0 \ , \tag{1}$$

where $x(t)$ is the amount of a substance in a volume (here the $CO_2$ concentration in the dry air) and $u(t)$ is the exogenous input/output flow of the substance (here $CO_2$ intake by fossil fuel combustion) which is independent of $x$. Equation (1) is of the same type of the key equation (15) (Section 2) of this paper.

Several detailed packages have been developed within the research studies of the global climate prediction. Here we will derive and employ a simple dynamic model mimicking the chemical kinetics of the $CO_2$ exchange between atmosphere and soil biomass and atmosphere and ocean. The model is described in the Appendix A as a set of state equations, where each state variable expresses the $CO_2$ amount of different Earth's reservoirs (as such or chemically modified forms), included the three main fossil fuel deposits (coal, oil and natural gas). Model simplification under reasonable assumptions leads to a first order differential equation with a pair of unknown parameters to be fitted on the historical data of $CO_2$ concentration and fossil fuel emissions. The pair of parameters accounts for

(1) the equilibrium concentration of the atmospheric $CO_2$, which is found close to the ice-core mean value of the last two thousand years of the Holocene,

(2) the land/ocean absorption time constant which is found to be close to half a century.

Raw data uncertainty reflects onto the parameter estimate variance which has been estimated to accompany predictions with statistical bounds. The advantage of simple physical models lies in that their parameters can be easily related to historical data and model structure can be checked and optimized with the help of statistical tests. As a such they may be a valuable tool for checking and pruning more detailed models.

In Section 3, historical data of the fossil fuel depletion are extrapolated with the help of the Meixner curve [25], a typical logistic curve. The aim is to predict the future depletion constrained by current reserves. A prediction until zero depletion has been already reported in [7]. Here it will be revised by exploiting a better estimates of the current reserves. The predicted depletion, converted into equivalent $CO_2$ emission, will become the input to the previous dynamic model. Starting from the current epoch, the model will integrate the extrapolated emissions decremented for land and ocean absorption, thus providing the finite-reserve projection of the atmospheric $CO_2$ concentration until 2200, only accounting for fossil fuel emissions. The projection is accompanied by an uncertainty band as a result of the employed experimental data. We should stress the fact that the CO2 concentration projection is just the result of experimental/literature data and elementary models, thus favoring robustness and easing interpretation and updating.

Comparison with projections of complex simulation packages driven by business-as-usual and mitigation policies is not immediate. Projections are included in the IPCC sixth assessment report [34] but the relevant data files seem unavailable. In a previous paper [7], we compared the projections of $CO_2$ emissions, with finite-reserve projections and with the current policy ones.

Other human activities like deforestation, generically land-use change, (see Figure 2), though possible sources of $CO_2$ and other GHG emissions, will be neglected in the paper since their emissions, not deriving from a resource reservoir, cannot be forecast with the method explained in the paper. A further interesting fact, as pointed out in [26], is the correlation of the atmospheric $CO_2$ concentration rate with temperature records of the Pacific Ocean El Niño phenomenon. Correlation might be modeled as a temperature dependence of the kinetic constant $k=\tau^{-1}$ of the key equation (14) (see also (1)), which scales the $CO_2$ exchange between ocean and atmosphere, but unfortunately El Niño historical data can only be extrapolated over short times (a few years), thus preventing their use in a long-term forecast model.

## 2 A simple dynamic model of CO₂ absorption by land/ocean sinks

### 2.1 Introduction

The aim of the section is to formulate a dynamic model of the annual mean carbon cycle, which excludes the seasonal perturbation (which average to zero), but is capable of fitting the atmospheric $CO_2$ drift of the industrial era. Fossil fuel ~~and~~ other emissions and their absorption by land and ocean sinks will be accounted for. To this end, it seems natural to formulate atmosphere ( $s=1$ ), ocean ( $s=2$ ), land (better soil, $s=3$ ), cement constructions ( $s=4$ ) and fossil fuel deposits ( $s=5$ ) as separate *reservoirs* capable of storing an amount $x_s$ of $CO_2$ under different forms which are subjected to $CO_2$ emission and absorption as driven by physical laws and human activities. Something like this has been sketched by D. Archer in [23], Chapter 4, and referred to as the 'Carbon Cycle orrery' (see also [31] and [32]). The reservoir state variables (briefly states or levels) are collected into the column vector $x=[x_1, x_2, x_3, x_4, x_5]$, where the inline notation from [11] has been adopted. The term reservoir implies both emission and uptake, whereas sink just indicates uptake, and fossil fuel deposit just emission.

Here we propose a dynamic system formulation [11], where the $CO_2$ amount $x_s(t)$ or a chemical compound containing it, like $CO_3^{2-}$, of each reservoir is a state variable, whose time derivative equals a (linear) combination of input (positive) and output (negative) flows $\pm v_{sh}(x)$, which depend on the reservoir levels and on the flows $\pm u_s$ generated by human activities. The detailed system of equations is derived in the Appendix A and then simplified under reasonable assumptions into a pair of state equations (first order differential equations) driven by the emission $u_5$ of fuel combustion. The first equation indexed by $s=1$ to be developed in this section expresses the annual mean concentration $x=x_1=[CO_2]_{atm}$ of the atmospheric $CO_2$ as the combination of anthropogenic emissions and of the land/ocean absorption (the subscript 1 will be dropped for simplicity's sake).

The second equation to be employed in Section 3 expresses the depletion of the fossil fuel reserves $r=x_5$. which will be employed to constrain the future $CO_2$ emissions. Emission extrapolation until the zero-reserve date is approached, once entered the atmospheric reservoir equation of Section 2, will allow us to predict accumulation of the relevant $CO_2$ concentration.

In the next sub-sections, the first equation will be directly derived by physical/chemical reasoning thus providing a justification of the dynamic model of Appendix A. The resulting equation will be converted into a perturbation equation around the unknown atmospheric $CO_2$ equilibrium $\bar{x}=\bar{x}_1$ and employed for estimating equation parameters (equilibrium and time constant) from the historical data listed in the Appendix 2.

### 2.2 Annual mean rate and concentration: definition, measurements and units

The annual mean concentration $\bar{x}(t)$ of the atmospheric $CO_2$ is defined by the integral

$$\bar{x}(t)=\frac{1}{T}\int_{t-T/2}^{t+T/2} x(\tau)d\tau \text{ [ppm]}, \ T=1\text{ y} \ , \qquad (2)$$

where $x(t)$ denotes the current concentration in the dry atmosphere measured in part per million of mole [ppm] and the time $t$ is given in fractions of year [y]. Thus, any zero-mean component of $x(t)$ in the integration interval does not contribute to $\bar{x}(t)$.
The integer $t_0=\text{floor}(t)$ (say $t_0=1750$) corresponds to time 0:0 of January 1$^{st}$, 1750, and the current year is denoted by $t_i=t_0+iT=t_0+i, i=0,1,...$ , where $t_0$ must be chosen. The generic time instant is defined as $t=T_0+iT+\tau, 0\leq\tau<T=1$. Since the mean $\bar{x}_1$ is measured from the start of January to the end of December, the corresponding sample $\bar{x}(i)$ must be referred to the solar year mid time $s_i=t_i+T/2=t_i+1/2$, that is

$$\bar{x}(i)=\bar{x}(s_i)=\bar{x}(t_i+1/2) \ . \qquad (3)$$

The annual mean concentration rate $\dot{\bar{x}}(t)$, centered in $t$, can be proved to coincide with the following increment

$$\dot{\bar{x}}(t)=T^{-1}(x(t+T/2)-x(t-T/2)) \text{ [ppm/y]} \ . \qquad (4)$$

Thus, the mean rate $\dot{\bar{x}}(i)=\dot{\bar{x}}(s_i)$ of the variable $x$ in the year $t_i$ is defined in agreement with (4) and [15] as the *averaged* concentration between the end of December and the start of January, and referred to the mid year date $s_i=t_i+1/2$.

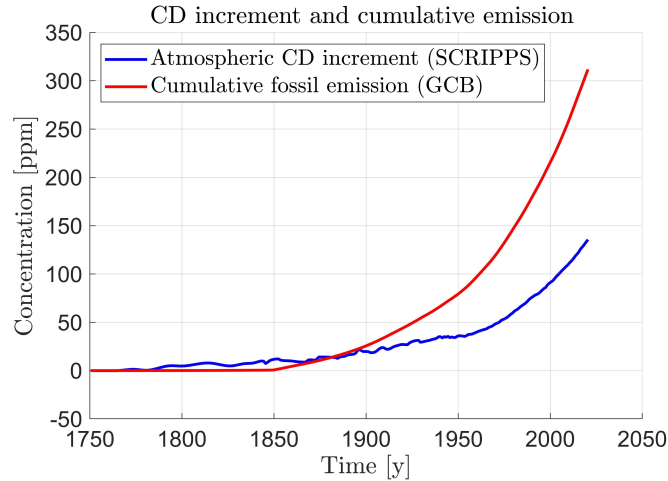

Figure 1 - Mean annual atmospheric $CO_2$ concentration increment since 1750, from SCRIPPS data, and the cumulative GCB fossil fuel emission.

Figure 1 shows the measurements of the mean $CO_2$ concentration increment $\breve{X}(i)=\breve{x}(i)-\breve{x}(0)$ (blue color, the 'breve' mark denotes measurements) from the SCRIPPS data record [17] since $s_0=1750.5$, and the measured cumulative sum $\check{C}(i)=T\sum_{k=0}^{i}\check{c}(t_k)$ [ppm] (red color) of the fossil fuel emissions [ppm/y] from GCB data. The simplified notation $c=u_5$ will be used throughout. All the measurements are in concentration [ppm] and concentration rate [ppm /y] units.

The natural measuring unit of the $CO_2$ amount (the state variables $x_s$) in the reservoir $s$ would be a mass unit like metric gigatonnes [$GtCO_2$], but we prefer the unit of the $CO_2$ concentration in the dry atmosphere, expressed in part per million of mole [ppm]. The conversion factor $\mu_{CO2}$ from $CO_2$ mass to concentration holds

$$\mu_{CO2}=\frac{1}{m_{ppm}}=\frac{1}{7.819}\frac{\text{ppm}}{\text{GtCO}_2} \quad , \tag{5}$$

where $m_{ppm}$ denotes the mass of one part per million of $CO_2$ molecules in the dry atmosphere. The proof comes by recalling [8] that the mass $m_{atm}$ of the Earth's atmosphere amounts to about $m_{Earth} \simeq 5.148 \times 10^{18}$ kg and that the average molecular mass $M_{atm}$ of the dry air amounts $M_{atm}=0.02895$ kg/mol. The number $N_{atm}$ of air moles follows to be

$$N_{atm}=\frac{5.148\times 10^{18}}{0.02897}\text{ mol}=1.777\times 10^{20}\text{ mol} \quad . \tag{6}$$

As a result, one part per million (1 ppm) of $N_{atm}$ moles possessing the $CO_2$ molecular mass $m_{CO2}=0.044$ kg/mol has the total mass in (5), namely

$$m_{ppm}=10^{-6}N_{atm}m_{CO2}=10^{-6}\times 1.777\times 10^{20}\times 0.044=7.819\times 10^{12}\text{ kg}=7.819\text{ GtCO}_2 \quad . \tag{7}$$

## 2.3 Atmospheric $CO_2$ state equation derivation

$CO_2$ is constantly exchanged among atmosphere, biomass, soil and ocean [1]. A big deal of $CO_2$ is taken out from atmosphere by photosynthesis (around 120 $GtCO_2$/year), but at the same time half of this goes back to air by plant emission during night and the daily animal breathing. Part of the remaining goes to soil after the plant death and, due to bacterial fermentation, again into air, part is washed away as $CO_3^{2-}$ by surface fresh water into sea. An estimate of the net exchange can be performed by computerized models. The results of different models have been recently collected and averaged by the Global Carbon Budget project (see [4] and [12]) and graphically plotted in Figure 2.

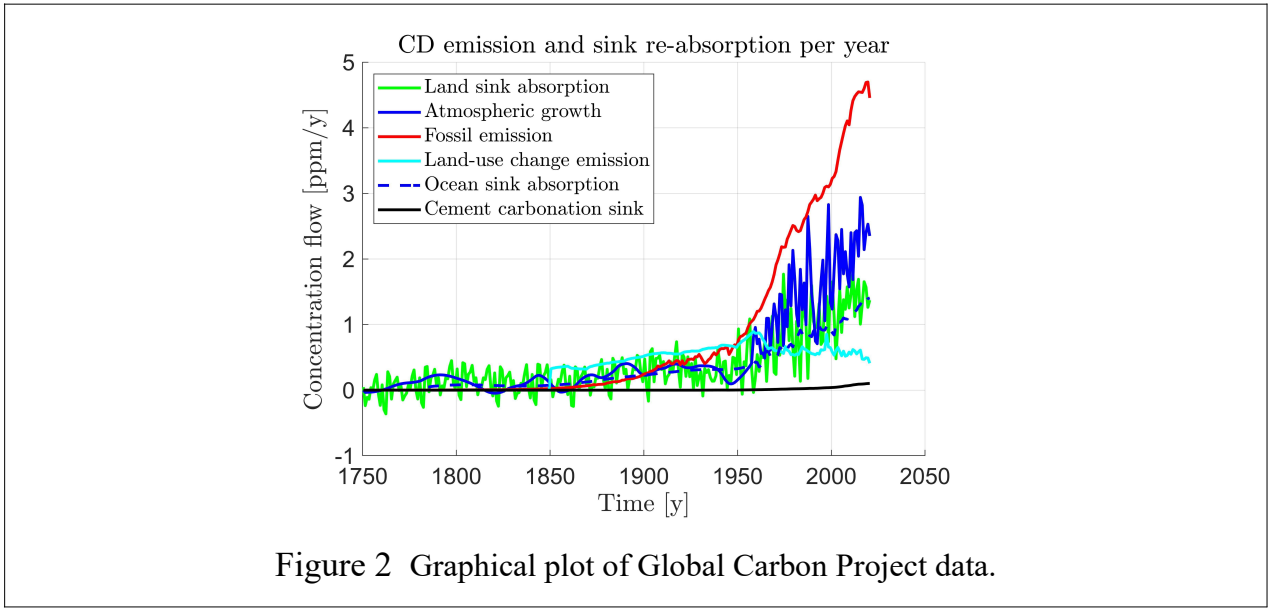

Figure 2  Graphical plot of Global Carbon Project data.

As seen in Figure 2, the main reservoirs re-absorbing the atmospheric $CO_2$ fossil emissions are the so-called ocean and land sinks. The relevant raw data comes from [4] and are summarized in the Appendix 2 together with their uncertainty. The $CO_2$ reservoir in seawater can be explained by the seawater reactivity, which, by its own nature, is alkaline in character. The equilibrium constant of the relevant reaction

$$CO_2(gas) + H_2O \Leftrightarrow CO_2(aq) + H_2O \Leftrightarrow H_2CO_3 \qquad (8)$$

has been discussed in [3], [10], [23] and [24]

The land sink can be mainly explained by the photosynthesis, which encompasses, in the very first stages, a reaction similar to (8) between carbon dioxide and the water which constitutes the cell cytoplasm. $CO_2$ enters the cell through the cell membrane, where it is incorporated into already existing organic carbon compounds, such as ribulose bisphosphate (RuBP). Using the organic compounds ATP (adenosine triphosphate, a source of energy) and NADPH (nicotinamide adenine dinucleotide phosphate, a reducing agent), which are produced by light-dependent reactions, the resulting compounds are then reduced and removed to form further carbohydrates, such as glucose.

Being the reaction (8) the first common stage of both sinks, it can can be incorporated into the same mathematical treatment. We must consider reaction (8) from the kinetic point of view, as we are interested in situations where chemical reaction have not yet reached their equilibrium conditions. This happens because each year tons of $CO_2$ are emitted into the atmosphere by the combustion of fossil fuels, thereby disturbing the pre-industrial equilibrium. The accepted value for the pre-industrial equilibrium is around 280±10 ppm (part per million of mole) [23]. As in any kinetic-controlled reaction, we must consider the direct (from left to right) and inverse (from right to left) semi-reactions. However, the latter has a nearly constant rate, being the concentration [$H_2CO_3$] of carbonic acid in seawater and in cytoplasm considered constant, at least in the decade time span. By accounting for the direct reaction

$$CO_2(\text{gas}) + H_2O \Rightarrow H_2CO_3 \quad, \tag{9}$$

we can write a differential equation describing the atmospheric $CO_2$ depletion:

$$\text{direct reaction rate } v_{dir}(t) = d[CO_2]_{atm}(t)/dt = -k_{dir}[CO_2]_{atm}(t) \quad, \tag{10}$$

and a reverse reaction describing the land/ocean reservoir depletion:

$$\text{reverse reaction rate } v_{inv}(t) = d[H_2CO_3](t)/dt = k_{inv}[H_2CO_3](t) = \text{constant value} = \bar{v}_{inv} \tag{11}$$

The over bar in (11) denotes constancy. The total reaction rate results from the sum $v_{dir} + v_{inv}$, which holds:

$$d[CO_2]_{atm}(t)/dt = v_{inv}(t) + v_{dir}(t) = \bar{v}_{inv} - k_{dir}[CO_2]_{atm}(t) \quad. \tag{12}$$

This is only a part of the process, because every year a certain known amount of anthropogenic $CO_2$ is emitted into the atmosphere. Eq. (12), when completed by the corresponding term $[CO_2]_{anthropic}$, accounting for fossil fuel combustion, land-use change and cement production (see Figure 2), and the initial condition, becomes

$$\begin{aligned} d[CO_2]_{atm}(t)/dt &= \bar{v}_{inv} - k_{dir}[CO_2]_{atm}(t) + d[CO_2]_{anthropic}(t)/dt \\ [CO_2]_{atm}(t_0) &= [CO_2]_{atm,0} \end{aligned} \tag{13}$$

Equation (13) shows that the pre-industrial equilibrium $\underline{[CO_2]}_{atm}$ can be obtained by inserting $d[CO_2]_{atm}(t)/dt = d[CO_2]_{anthropic}(t)/dt = 0$, which provides

$$\underline{[CO_2]}_{atm} = \bar{v}_{inv}/k_{dir} \quad. \tag{14}$$

Notation simplification with the help of Appendix A and (14) allows (13) to be rewritten as

$$\begin{aligned} \dot{x}(t) &= -k(x(t) - \underline{x}) + u(t), \quad x(t_0) = x_{10} \\ x &= x_1 = [CO_2]_{atm}, u = d[CO_2]_{anthropic}(t)/dt, k = k_{dir} \end{aligned} \tag{15}$$

Eq. (15) depends on the unknown parameters $\{\underline{x}[\text{ppm}], k[1/s]\}$ to be estimated from historical data. For the scope of this paper, (15) should be rewritten in terms of the mean annual concentration $\bar{x}(t)$ defined in (2) and the mean annual rate $\bar{u}$ of $[CO_2]_{anthropic}(t)$, which from (4) holds

$$\bar{u}(t) = T^{-1}([CO_2]_{anthropic}(t+T/2) - [CO_2]_{anthropic}(t-T/2)) \quad. \tag{16}$$

Because of linearity - integration and derivative commute - (15) does not change in terms of the mean variables, which suggests to keep the simpler notations of (15), thus posing $x=\bar{x}$ and $u=\bar{u}$. A similar equation but referred to the ocean reservoir was written by Revelle in [39].

## 2.4 Discretization and parameter estimation

In order to employ the measurements $\{\check{x}(i), \check{u}(i)\}, i=0,1,\ldots,N-1$ of the mean values $x=\bar{x}$ and $u=\bar{u}$, which refers to the mean year date $s_i = t_i + 1/2$, equation (15) must be integrated in the time interval $S(i)=\{t; s_i \leq t < s_{i+1}\}$. Integration is straightforward because of linearity and provides:

$$x(i+1) = \underline{x} + a(x(i) - \underline{x}) + \int_{s_i}^{s_{i+1}} \exp(-k(t_{i+1} - \tau)) u(\tau) d\tau \Rightarrow$$
$$\Rightarrow \Delta x(i) = x(i+1) - x(i) \simeq -(1-a)(x(i) - \underline{x}) + b(a) u(i), \quad (17)$$
$$a = \exp(-kT), \quad b(a) = \frac{1 - \exp(-kT)}{k} = \frac{(1-a)T}{\log(1/a)}$$

The approximation in the second row of (17) becomes acceptable if $kT \ll 1$ and $|1 - u^{-1}(i) u(\tau)| \ll 1$, leading to a fractional error well less than 0.01 thus absorbed by the measurement errors. By replacing $x$ and $u$ with their measurements, Eq. (17) converts into the following regression equation (referred to as *differential regression*):

$$\Delta \check{x}(s_i) = -(1-a)(\check{x}(i) - \underline{x}) + b \check{u}(i) + \Delta \tilde{x}(i), i=0,1,\ldots,N-1, \quad (18)$$

with the pair $\{a, \underline{x}\}$ of unknowns and the measurement error $\Delta \tilde{x}$. Under the assumption $kT \ll 1, T=1$, (18) can be approximated by the linear regression equation

$$\Delta \check{x}(i) \simeq -k(\check{x}(i) - \underline{x}) + \check{u}(i) + \Delta \tilde{x}(i), i=0,1,\ldots,N-1, \quad (19)$$

with unknowns $\{k, \underline{x}\}$. In principle, the input measurement $\check{u}(i)$ should include the main anthropogenic emissions, namely the fossil fuel emission $c=u_5$ and the land-use change emission $u_3$ as reported in Figure 2 and in Table 5. Here only $c=u_5$ will be included, since, unlike $u_3$, it can be extrapolated as done in Section 3, the aim being to a long-term forecast of the atmospheric $CO_2$ concentration.

For completeness, the parameter estimates $\{\hat{\underline{x}}, \hat{a} = \exp(-\hat{k}T)\}$ of the differential regression (18) will be checked from the *integral regression* equation

$$\check{x}(i) = \underline{x} + a^i(\check{x}(0) - \underline{x}) + b(a) \sum_{k=1}^{i} a^{i-k} \check{u}(k-1) + \tilde{x}(i), \quad (20)$$

where the annual mean $\check{x}(i)$ of the Mauna Loa data equals the discrete-time integration of the first equation in (17).

Under the assumption of a statistically independent, zero-mean and stationary error $\tilde{x}(i)$,

of a zero-mean and non-stationary emission rate error $\tilde{u}(i)$ and of

$$1 - a \simeq kT \ll 1, \quad b(a) \simeq T = 1 \quad , \tag{21}$$

the measurement error $\Delta\tilde{x}(i)$ and its (a priori) variance can be approximated with the help of Table 5 as follows (let us recall the notation $c = u_5$ ):

$$\begin{aligned}&\Delta\tilde{x}(i) \simeq \tilde{x}(i+1) - \tilde{x}(i) - T\tilde{u}(i), \quad T = 1 \text{ y}\\&\text{var } \Delta\tilde{x}(i) \simeq 2\,\text{var}\,\tilde{x} + T^2 \text{var}\,\tilde{u}(i) = 2\sigma_1^2 + (T\rho_c\check{c}(i))^2 < 0.1\\&\sigma_1 \simeq 0.12\ \text{ppm}, \rho_c \simeq 0.05, |T\check{c}(i)| < 5\ \text{ppm}\end{aligned} \tag{22}$$

Values in (22) come from GCB [16].
By denoting the estimated parameters with $\hat{\underline{x}}, \hat{a}$ , the *differential residuals* are defined by

$$\Delta\hat{\tilde{x}}(i) = \Delta\check{x}(i) + (1-\hat{a})(\check{x}(i) - \hat{\underline{x}}) - b(\hat{a})\check{u}(i), \quad i = 0,.., N-1 \quad , \tag{23}$$

whereas the *integral residuals* are defined by

$$\hat{\tilde{x}}(i) = \check{x}(i) - \hat{\underline{x}} - \hat{a}^i(\check{x}(0) - \hat{\underline{x}}) - b(\hat{a})\sum_{k=1}^{i}\hat{a}^{i-k}\check{u}(k-1) \tag{24}$$

The residuals in (24), obtained by integrating the first equation in (17) with the parameter estimates of the differential regression (18) are referred to as *cumulative residuals*. The differential residual RMS (playing the role of the a posteriori standard deviation) is denoted by $\hat{\tilde{\sigma}}_{\Delta x} = \sqrt{N^{-1}\sum_{k=0}^{N-1}\Delta\hat{\tilde{x}}(k)^2}$ . A similar expression applies to other residuals.

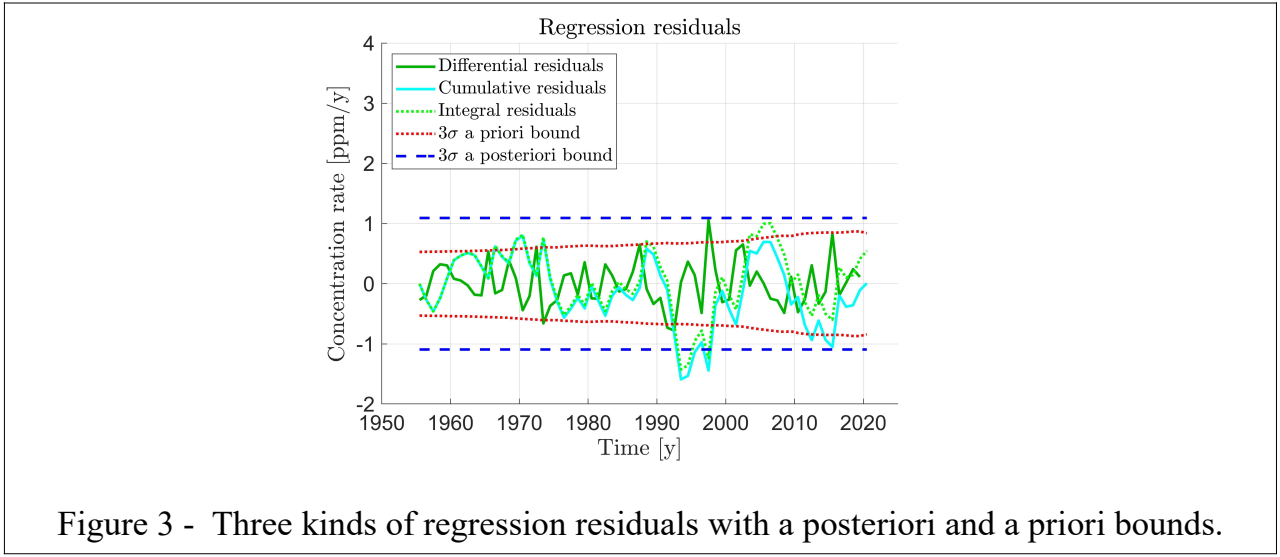

Figure 3 - Three kinds of regression residuals with a posteriori and a priori bounds.

Figure 3 shows the differential residuals $\Delta\hat{\tilde{x}}(i)$ in (23) (dark green) for $t_0 = 1955$ and the $3\sigma$ a priori bound from (22) compared with the a posteriori bound of the residual

RMS. The a priori bound increases because of the non-stationary emission error $\widetilde{u}$ in (22). As expected, cumulative (cyan color) and integral residuals (point-wise green) track each other with a small drift due to slightly different parameter estimates (see Table 1). The integral residuals tend to be larger (together with their RMS in Table 1) than differential ones, although both are zero mean, because of integrated mid frequency components. For instance, the negative overshoot of the integral and cumulative residuals since 1990 is not fortuitous, but is mainly due to Pinatubo volcanic eruption [26], which forced a short-term decrease of the $CO_2$ growth rate.

Estimation of the parameter covariance matrix employs the linear regression equation (20), which is rewritten in the following vectorial form:

$$\begin{aligned}\breve{y}=\Delta\breve{x}-\breve{u}=U\,a+\widetilde{y},\ \widetilde{y}(i)=\Delta\widetilde{x}(i), E\{\widetilde{y}\widetilde{y}^T\}=\widetilde{S}^2=\mathrm{diag}(\mathrm{var}\,\Delta\widetilde{x}(i))\\ a=[a_0=k\,T\,\underline{x},a_1=k\,T],\ U=[\boldsymbol{u}_0\ -\breve{x}],u_0(i)=1\end{aligned} \quad (25)$$

The a posteriori covariance matrix $\hat{S}_a^2$ and the parameter standard deviations $\hat{\sigma}_x,\hat{\sigma}_\tau$ of $\hat{\bar{x}},\hat{\tau}=\hat{k}^{-1}$ hold

$$\begin{aligned}\hat{S}_a^2=\hat{\widetilde{\sigma}}_{\Delta x}^2(U^T U)^{-1}\\ \hat{\sigma}_x=\hat{a}_1^{-1}\sqrt{\hat{S}_a^2(1,1)}\ [\mathrm{ppm}],\ \hat{\sigma}_\tau=T\,\hat{a}_1^{-2}\sqrt{\hat{S}_a^2(2,2)}\ [\mathrm{y}]\end{aligned}, \quad (26)$$

where $\tau$ is the time constant of $k$.

## 2.5 Regression results

Differential and integral regression results are shown in Table 1 for $t_0=1955$. Annual mean data are available by SCRIPPS program [17] starting from $t_0=1750$ and before, but $CO_2$ rate data from ice cores (before 1959) look rather irregular as shown in Figure 4, which suggests to restrict elaboration starting from $t_0=1955$ (the shaded area in Figure 4).

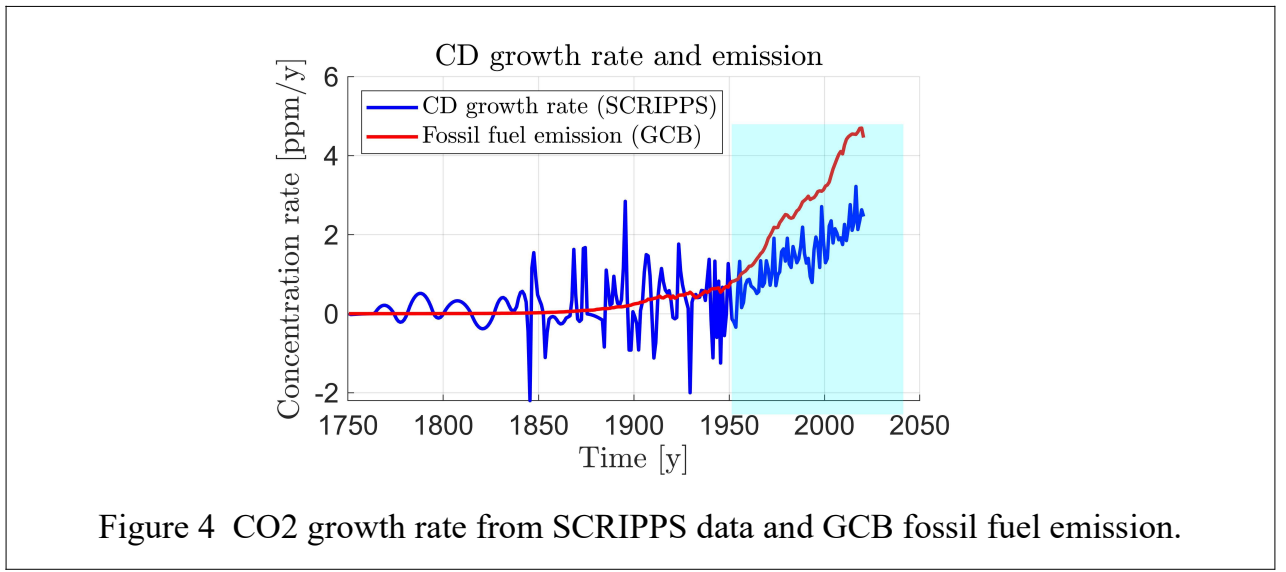

Figure 4  CO2 growth rate from SCRIPPS data and GCB fossil fuel emission.

Regression results are summarized in Table 1. It should be stressed that the estimated time constant $\hat{\tau}$ (lifetime in the carbon cycle and climate change literature) refers to the absorption by the whole Earth sinks, mainly land and ocean, of the annual mean atmospheric CO2 concentration (the annual periodic two-way exchange has not been accounted for by the previous dynamic models). In other terms, all the absorption kinetic constants are summed up as in the entry (1,1) of the matrix $A$ in (45). Literature estimates are rather sparse. Revelle [39] estimates for the ocean sink 10 years. Archer [23] computes 100 years. IPCC Working Group I reports in [41] from 5 to 200 years, by remarking that 'No single lifetime can be defined for $CO_2$ because of the different rates of uptake by different removal processes'.

| Table 1 Differential and integral (within brackets) regression results (from 1955) ||||||
|---|---|---|---|---|---|---|
| No | Parameter | Symbol | Unit | Estimated value | A posteriori standard deviation | Comments |
| Integral regression estimates are in brackets. A posteriori standard deviation comes from (26). ||||||
| 1 | Kinetic constant | $\hat{k}$ | 1/y | 0.0188 (0.0191) | 0.0015 | |
| 2 | Time constant | $\hat{\tau}$ | y | 53.1 (52.3) | 4.3 | |
| 3 | Equilibrium concentration | $\underline{\hat{x}}$ | ppm | 285.9 (286.4) | 4.0 | |
| 4 | Historical equilibrium | $\underline{x}_{hist}$ | ppm | 279.6 | 3.0 | until 1850 |
| 5 | Residual RMS | | ppm | 0.364 (0.540) | NA | |

Figure 5, left, shows the measured atmospheric $CO_2$ rate $\Delta \check{x}(i)$ (blue color) since 1955, the relevant estimate (dashed red) and the estimated residuals $\Delta \widetilde{\hat{x}}(i)$. The residual short-term fluctuations would be partly explained by including in (18) as a scaled input the temperature anomaly of one of the Pacific Ocean equatorial belts, monitoring the El Niño phenomenon [26]. The differential residual RMS would reduce to below 0.3 ppm (-20%), but historical data cannot be extrapolated thus preventing, as already remarked, their use in Section 3. Figure 5, right, shows the raw mean $CO_2$ concentration $\check{x}(i) - \underline{\hat{x}}$ (blue line) and the estimate (dashed red) with respect to the estimated equilibrium (the dashed zero line). Regression residuals are also shown.

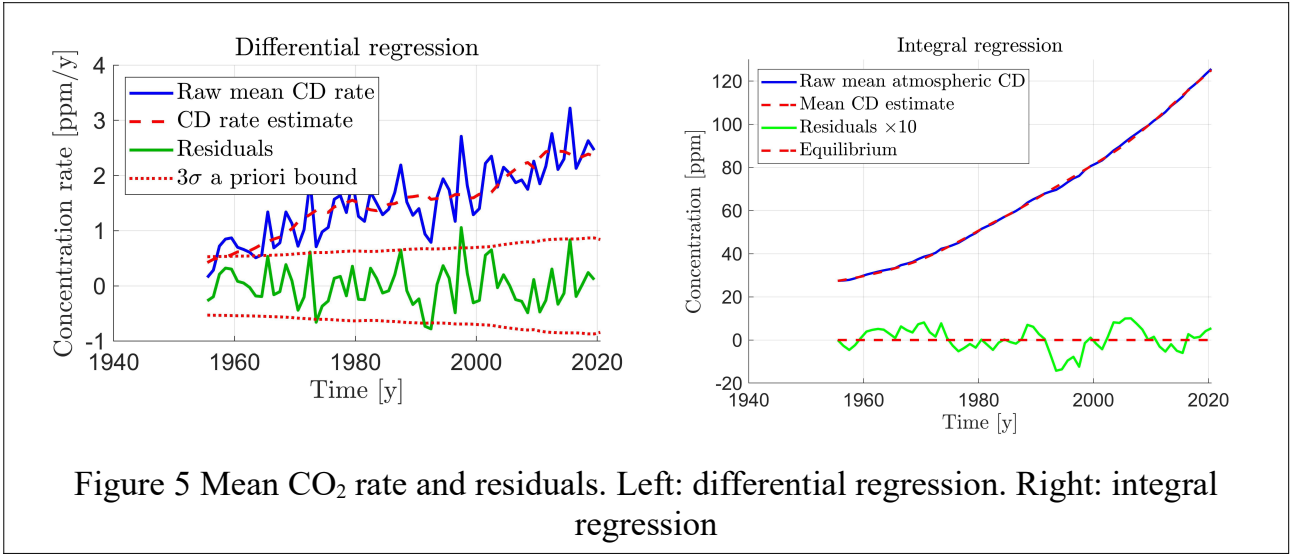

Figure 5 Mean $CO_2$ rate and residuals. Left: differential regression. Right: integral regression

## 2.6 Regression extension to whole industrial era

The regression restriction to the recent epoch since 1955 may rise some question about estimate significance and robustness. Extension to industrial era is deemed not necessary for future predictions, since as Figure 4 shows, significant increment of fossil fuel emissions (especially of oil and natural gas) just started around 1950. Aiming to prove regression robustness, though in presence of irregular data until 1950, a sequence of $M=6$ differential regressions has been done from $1860$ until $1960$, with incipit dates equal to $t_0(k)=1800+(k-1)20, k=0,1,...,M-1$.

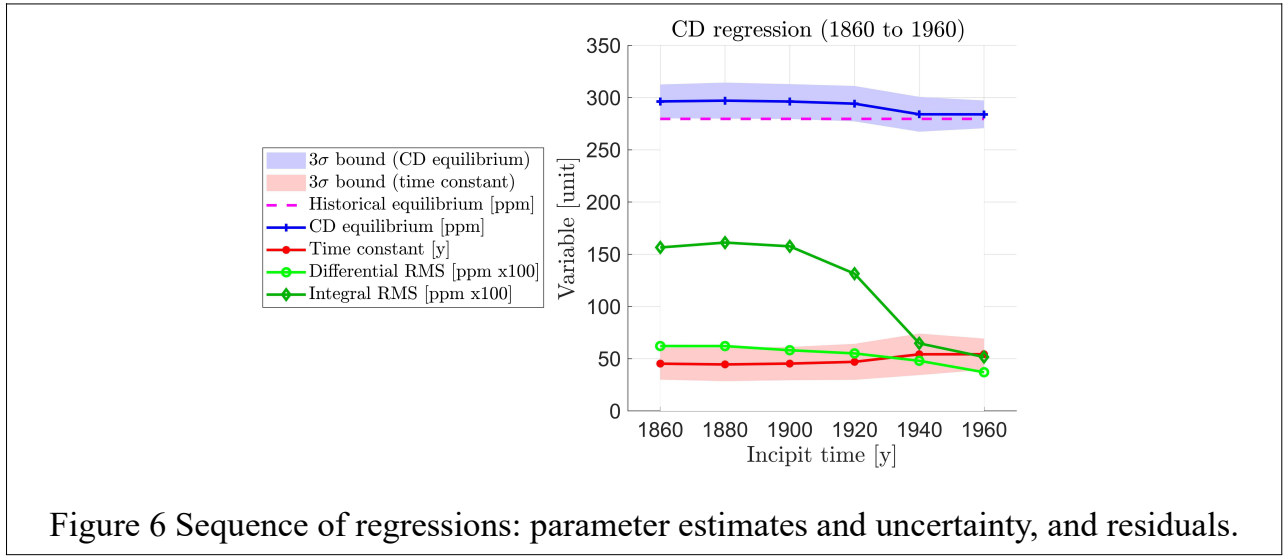

Figure 6 Sequence of regressions: parameter estimates and uncertainty, and residuals.

Figure 6 shows the sequence of parameter estimates (blue and red lines) and their $3\sigma$ uncertainty band. The band width increases toward earlier dates but is partly compensated by a larger size of the measured samples. What is more significant is the bias of the $CO_2$ equilibrium $\hat{x}$, accompanied by the explosion of the integral residuals (dark green). This occur because of the integration of low-frequency differential residual components, and implies that the dynamic model (15), as it is (and the relevant emission data) cannot accurately explain the low-frequency components of the atmospheric $CO_2$ concentration before 1950. The issue has not been further investigated, but apart data irregularity, a possible explanation as mentioned in [23] is that the main source of $CO_2$ emission in the XIX century and beyond was deforestation.

## 3 Extrapolation of the future $CO_2$ concentration

### 3.1 Introduction and scope

The scope of this section is to employ Eq. (15) and the estimated parameters in Table 1 to extrapolate the atmospheric concentration of $CO_2$ from the present, in other words from $t_N=2020$, the last year of the data employed in Section 2. To this end, we need to forecast the input signal $u(t), t > t_N$, which, as already said, will be limited to fossil fuel emission $c(t)$. Forecasting will be done by extending past fuel consumption $\check{c}(t_i), t_0 \leq t_i \leq t_N$ (converted into $CO_2$ emissions) by means of a parameterized analytic model, that is the skewed Meixner distribution (MD, or skewed/generalized hyperbolic secant distribution, from J. Meixner, 1908-1994) (see [7] and [25]). Forecasting will be constrained by the available reserves, that is by the estimated amount $\hat{r}(t_N)$ of the fossil fuel deposit at the present date $t_N=t_P$, whose equation from Appendix A holds:

$$\dot{r}(t) = -c(t), \ r(t_N=t_P) = r_P \ . \tag{27}$$

To do this, fossil fuel consumption and reserves must be split into the three categories $f=c,o,g$ of fossil fuels as accounted for by historical data, namely coal $(f=c)$, oil $(f=o)$ and natural gas (methane) $(f=g)$. The total extrapolated $CO_2$ emission

$$\hat{c}(t_i) = \sum_{f=1}^{3} \hat{c}_f(t_i), i=N, N+1, \ldots, End=N+M-1 \tag{28}$$

where $t_{End}=t_{N+M-1}$ is the estimated zero-reserve date, will be employed in (15) to in turn predict the $CO_2$ concentration $\hat{x}(t_i)$ in the atmospheric reservoir.

### 3.2 Reserves and resources

In order to quantify the amount of fossil fuels left for use by humankind, let us distinguish between reserves and resources. Resource is that amount of a natural commodity (in this case fossil fuels) that exists in both discovered and undiscovered deposits. Reserves are that subgroup of a resource that have been discovered, have a known size, and can be technically recovered at a cost that is financially feasible at the present price of that feed stock. Hence reserves will change with the price, unlike resources, which include the full amount of that stuff that can be technically recovered at any price. Factors that affect profitability include the demand, market price, extraction/transport costs, new technologies, and so on. As a consequence, the known reserves of fossil fuels vary in time, with an increasing trend in the last decades, as shown in Figure 7 (data from OWID, [13]).

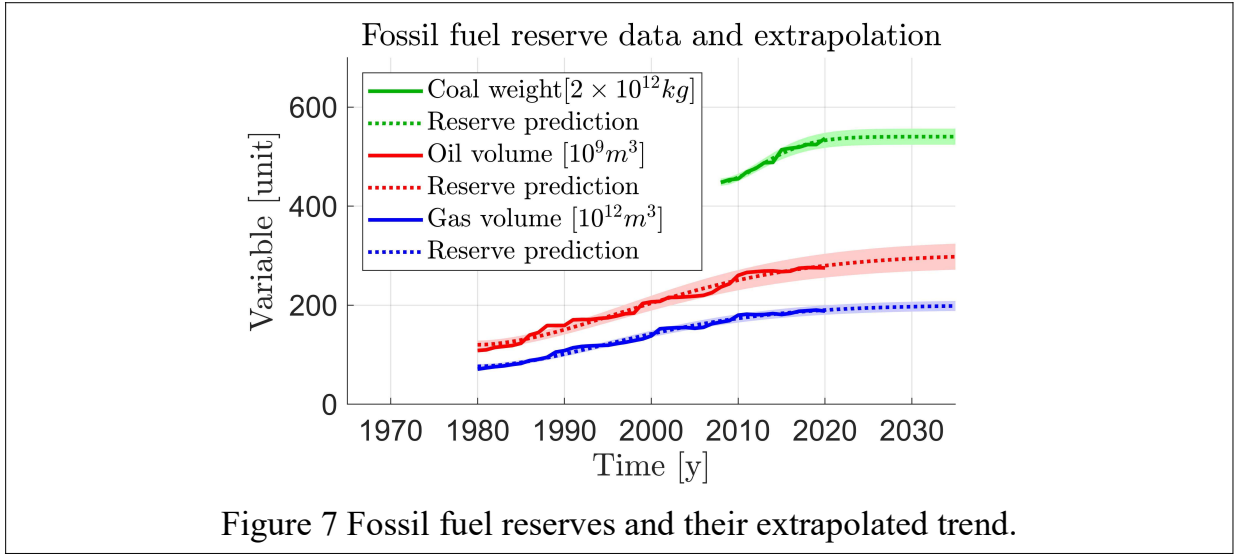

Figure 7 Fossil fuel reserves and their extrapolated trend.

Let us denote the reserve amount of a generic fossil fuel with $r(t)$. By supposing that the reserve trend of the raw data in Figure 7 will attain a constant value $r_\infty$ in the years to come, the trend can be extrapolated if we know the diminishing law of the future marginal reserves $m(t) = r_\infty - r(t)$ with initial condition $m_0 = r_\infty - r_0$ at time $t = t_0$ (the raw data initial time). A law of this kind is usually arranged by assuming the relative variation $dm(t)/m(t)$ to be proportional (with negative sign) to the dimensionless time interval $ds(t) = dt/\tau$ as follows

$$\frac{dm(t)}{m(t)} = -n s(t)^{n-1} ds(t), \quad m(t_0) = m_0, \quad s(t) = \frac{t - t_0}{\tau} \quad . \tag{29}$$

In the case of $n=1$, proportionality becomes constant in time and the marginal reserves decrease exponentially to zero. Here we further assume $1 < n < 2$, which implies that the decrement is slightly faster than the exponential one, due to increasing difficulties in finding new reserves. Integration of (29) provides the explicit incremental law

$$r(t) = r_\infty - m_0 \exp(-s(t)^n), \quad t \geq t_0 \quad . \tag{30}$$

The asymptotic value $r_\infty$, which defines the ultimate reserve value, will be used for further analysis. The four parameters $\{\tau, m_0, r_\infty, n\}$ can be found by fitting the raw data in Figure 7 (solid lines) and depend on the fuel category $f$. Given the four parameter estimates, the reserve $r(t)$ in (30) can be extrapolated to future dates as in Figure 7 (dashed lines). The shaded bands around the extrapolated curve $\hat{r}(t)$ correspond to $3\sigma$ uncertainty. Value and estimated standard deviation $\sigma_{rf}$ of the ultimate reserves $r_{\infty f}, f = c, o, g$ are reported in Table 3.

Once assessed the reserves of the fossil fuels, we have to determine which amount of $CO_2$

will be emitted in their future combustion until depletion is reached. Chemically speaking, the mass of $CO_2$ produced by a unitary mass of fuel, denoted by $\mu_f, f=c,o,g$, can be inferred from the fuel chemical composition by balancing simple chemical reactions, like

$$\begin{aligned}&CH_4+2O_2 \Rightarrow CO_2+2H_2O \quad (\text{natural gas}, \mu_g=44/16=2.75\,\text{kg/kg}) \\ &2C_8H_{18}+25O_2 \Rightarrow 16CO_2+9H_2O \quad (\text{crude oil}, \mu_o=704/228=3.09\,\text{kg/kg}) \\ &C+2O_2 \Rightarrow CO_2 \quad (\text{coal}, \mu_c=44/12=3.67\,\text{kg/kg})\end{aligned} \qquad (31)$$

Natural gas is supposed to be methane (CH4), crude oil octane (C8H18) and coal simply C (carbon). However, since chemical composition of fuels, in particular that of coal and oil, is far from being expressed by a single chemical compound, we adopted another way to assess the conversion factor $\rho_f$ of each fuel type.

OWID database about fossil fuels [13] reports the annual consumption of each fuel type (oil, gas and coal), during the 1980-2020 period, and the relevant $CO_2$ emissions. Elaboration of these data has provided the estimate $\hat{\rho}_f$ of the conversion factor $\rho_f$ as the mean annual $CO_2$ emitted mass of the fuel unit mass, in the case of coal [kg/kg], and of the unit volume in the case of oil [tonne/m³] and natural gas [kg/m³]. together with T corresponding standard deviation $\sigma_{\rho f}$ has been also estimated. Results are collected in Table 2.

| Table 2 Conversion factors | | | | | |
|---|---|---|---|---|---|
| No | Fuel | $\mu_f$ [ kg/kg] from (31) | $\hat{\rho}_f$ (estimated) | Unit | $\sigma_{\rho f}$ |
| 1 | Coal | 3.67 | 1.92 | kg/kg | 0.04 |
| 2 | Oil | 3.09 | 2.28 | tonne/m³ | 0.12 |
| 3 | Natural gas | 2.75 | 1.87 | kg/m³ | 0.07 |

Table 3 reports, in the third column, the ultimate reserve values extrapolated from the raw data in Figure 7 with the help of (30), together with the standard deviation and the appropriate unit (fourth column). The ultimate reserve values expressed in equivalent $CO_2$ mass units (fifth column) follow from $\hat{R}_f = r_{\infty f} \hat{\rho}_f$. They can be converted into concentration units [ppm] by (5). The relevant standard deviation can be easily computed. The value is compared with fuel reserves estimated by a previous research [7]. Row 4 of the table shows, for completeness, the estimated reserves of the unconventional shale-oil. Shale oil originates from shale rock fragments by pyrolysis, hydrogenation or thermal dissolution. The largest reserves exist in the United States [14]. Shale oil reserves and consumption will not be treated in this paper.

Table 3 Predicted global fuel reserves and the estimated equivalent $CO_2$ emission in $GtCO_2$ ($Gt=10^9$ metric tonnes)

| No | Fuel | $r_{\infty,f}$ [unit] extrapolated from (30) | Unit | $\hat{R}_f$ [$GtCO_2$] | $R_f$ [$GtCO_2$] from [7] |
|---|---|---|---|---|---|
| 1 | Coal | $(1120\pm30)\times10^{12}$ | kg | $2150\pm95(1\sigma)$ | $2180\sim2850$ |
| 2 | Oil | $(303\pm8)\times10^{9}$ | $m^3$ | $690\pm54(1\sigma)$ | $639\sim787$ |
| 3 | Natural gas | $(200\pm3)\times10^{12}$ | $m^3$ | $373\pm19(1\sigma)$ | 362 |
| 4 | Shale oil ([14]) | $445\sim525\times10^9$ | $m^3$ | $1015\sim1200$ | Not applicable |

### 3.3 Fossil fuel emission extrapolation

Given the historical fossil-fuel emissions, we aim to predict the future $CO_2$ emissions by accounting for the constraint imposed by finite fossil-fuel reserves in Table 3. The method adopted is to predict the fuel emission of each fuel type (coal, oil, gas) by constraining the total future emission by the relevant fuel reserve.

Let us provide a simple formulation. Let us denote the cumulative consumption of the fossil fuel $f$ in equivalent emitted $CO_2$ mass [$GtCO_2$] by $C_f(t_i)=\sum_{k=N}^{N+i} c_f(t_k)$ and assume that the future consumption is bounded by the available reserve $R_f$, that is $C_f(t_{N+M-1})=R_f$. The future interval is defined by the end time of historical data $t_{N-1}$ and $t_Z>2100$ which will be chosen not smaller than the usual end time of the literature equal to 2100. The total (all the fuels) cumulative CD emission [$GtCO_2$] is denoted by $C(t_i)$ and the rate by $c(t_i)$ [$GtCO_2$/y]. As already said, the emission rate $c_f(t_i)$ of each fuel type is approximated and extrapolated by an analytic function $c_f(t)=g(t;\mathbf{p}_f)$ depending on the parameter vector $\mathbf{p}_f$ to be estimated from past data. The extrapolation curve to be used is the four-parameter skewed Meixner distribution, whose shape recalls an asymmetric bell as portrayed in Figure 8. It can be written as

$$g(t,\mathbf{p})=a\frac{\exp(\beta\sigma)}{\cosh(\sigma)}, \sigma=\frac{t-s}{\tau} \quad \mathbf{p}=[a,\beta,\tau,s] \qquad (32)$$

where $a=\max_t g(t)$ (the height of the maximum under $\beta=0$) is the scale factor, $s=\arg\max_t g(t)$ denotes location (the abscissa of the maximum under $\beta=0$), $\tau$ defines the width of the bell shape and $-1<\beta<1$ defines the skewness degree of the shape. It is not difficult to prove that the skewness degree $\beta$ becomes unidentifiable when measurements are restricted to a lobe of the bell (either left or right), that is either to $\sigma(t_k)<0$ or $\sigma(t_k)>0$. The former is our case, which forces to adopt the symmetric shape with $\beta=0$.

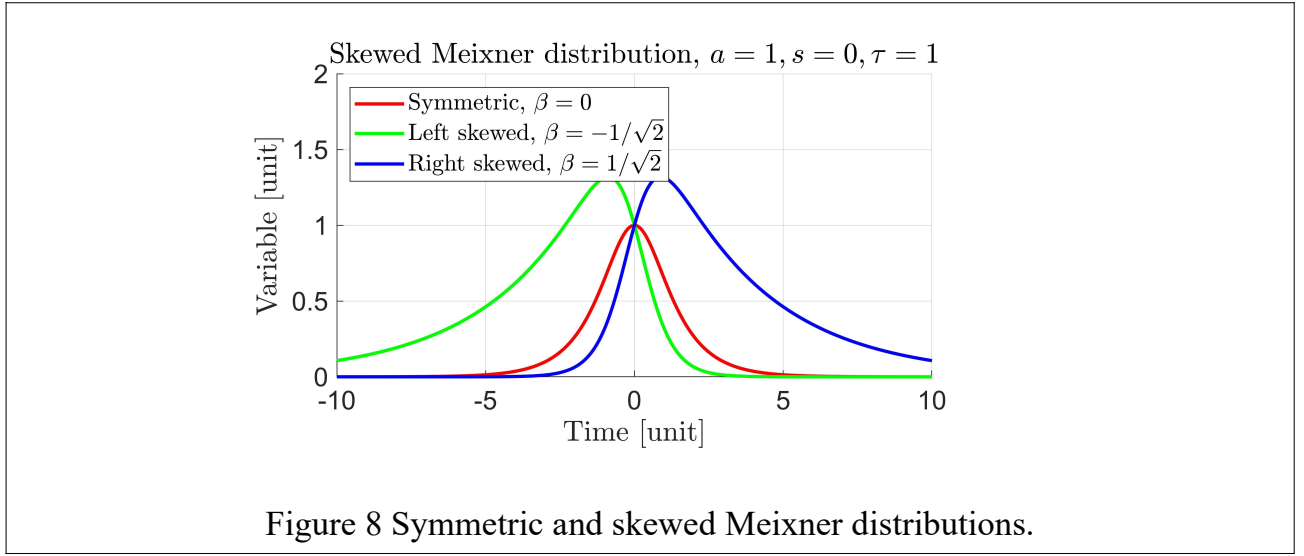

Figure 8 Symmetric and skewed Meixner distributions.

The nonlinear regression equations are the following:

$$\begin{aligned}\check{c}_f(t_k)&=g(t_k;\boldsymbol{p}_f)+\widetilde{c}_f(t_k), k=0,\dots,N-1, \quad \boldsymbol{p}_f=\{a_f,\tau_f,s_f\}\\ \hat{R}_f&=\sum_{k=N}^{N+M-1} g(t_k;\boldsymbol{p}_f)+\widetilde{R}_f\end{aligned} \quad (33)$$

The regression criterion $J$ to be minimized is the weighted square error

$$J(\boldsymbol{p}_f)=\frac{1}{N}\left(\sum_{k=0}^{N-1}\widetilde{c}_f^2(t_k,\boldsymbol{p}_f)w(t_k)+\widetilde{R}_f^2(\boldsymbol{p}_f)W\right). \quad (34)$$

where $w(t_k), W$ are positive weights accounting for uncertainty.
Figure 9 shows the extrapolated $CO_2$ emission profiles [GtCO2/y] by fuel, based on historical data from 1955 to 2020 (the irregular part of the mean profile, central solid line) and the minimization of the criterion in (34). The lower and upper bounds (pointwise and dashed lines) account for the uncertainty of the estimated parameter vector $\hat{\boldsymbol{p}}_f$ and of the estimated reserve $\hat{R}_f$ (see Table 3, column 5). As such, the uncertainty band around historical data is smaller than the extrapolation band, since the former is poorly affected by the reserve uncertainty.

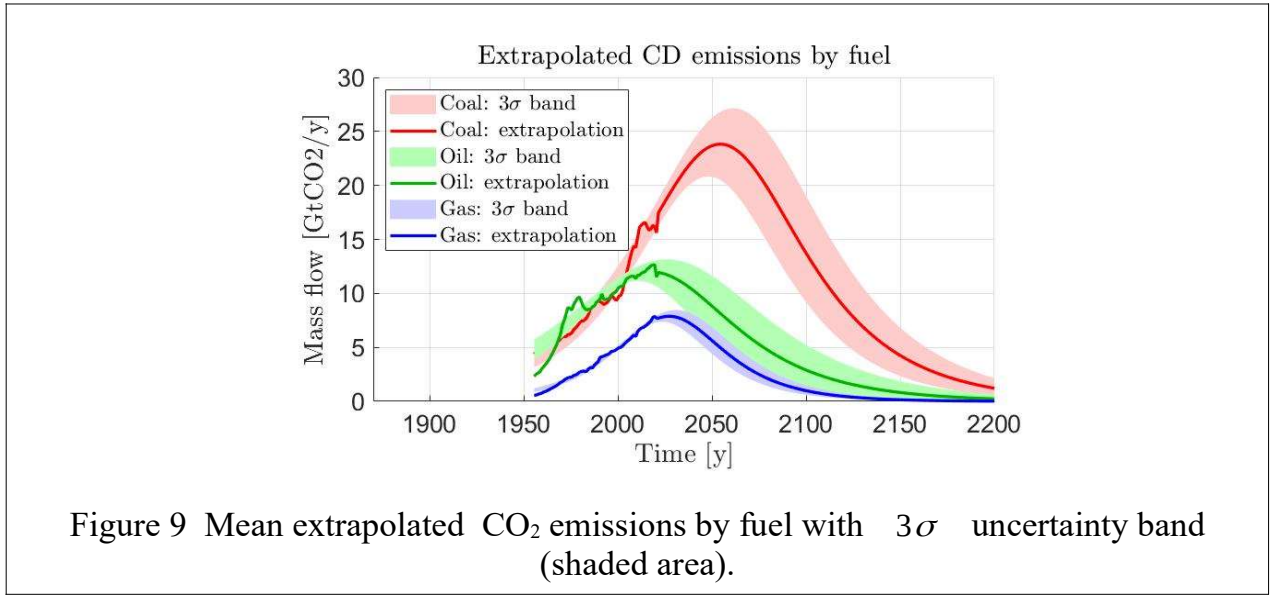

Figure 9 Mean extrapolated $CO_2$ emissions by fuel with $3\sigma$ uncertainty band (shaded area).

## 4  $CO_2$ uptake by sinks and emission by fossil fuel, the complete picture

### 4.1  $CO_2$ projections

The sum of the three emission profiles in Figure 9 is reported in Figure 10 together with the $3\sigma$ uncertainty band (shaded area).

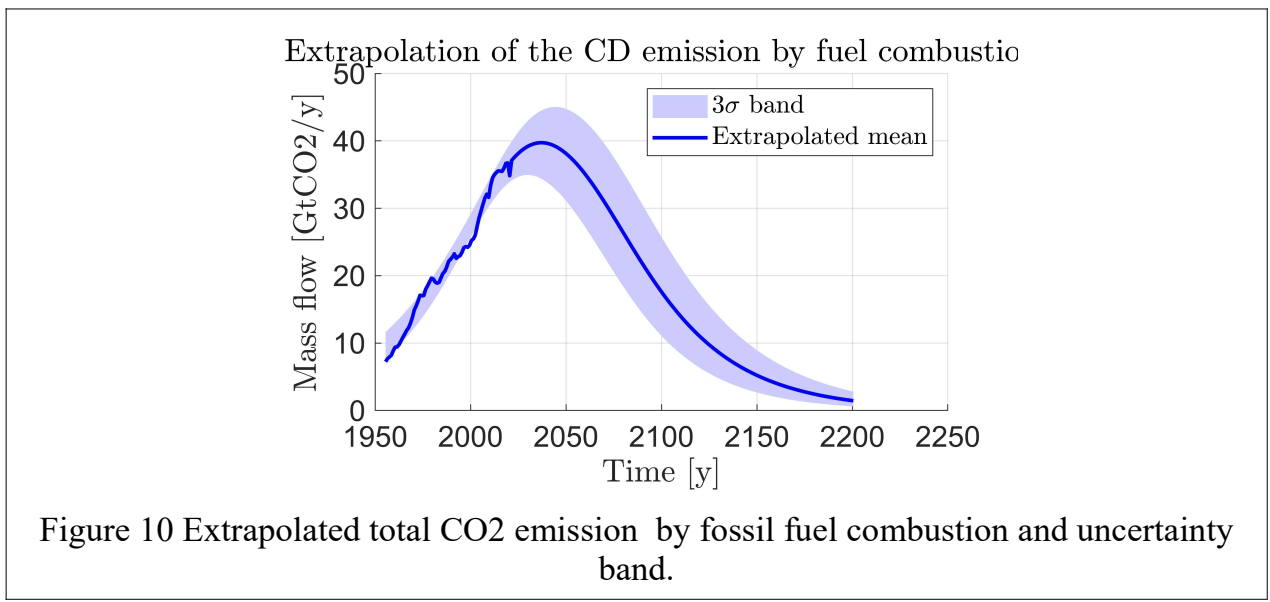

Figure 10 Extrapolated total CO2 emission by fossil fuel combustion and uncertainty band.

Replacement of the mean and $3\sigma$ lower and upper profiles in Eq. (15) as the input signal $u(t)$ and integration allows us to extrapolate until 2200 the (annual mean) atmospheric CO2 concentration as the opposite effect of the fossil fuel emission and of the land/ocean absorption. The mean extrapolation derives from the mean profile in Figure 10 and the estimated pair $\{\hat{\tau}, \hat{x}\}$ in Table 1, column 5. The $3\sigma$ lower and upper bounds derive from the $3\sigma$ profiles in Figure 10 and the $3\sigma$ parameter estimates $\{\hat{\tau}\pm 3\hat{\sigma}_\tau, \hat{\underline{x}}\pm 3\hat{\sigma}_x\}$, where the positive sign applies to the upper bound and the negative to the lower bound. The estimated standard deviations are reported in Table 1, column 5.

The resulting mean profile and the $3\sigma$ uncertainty band of the extrapolated $CO_2$ concentration are reported in Figure 11. The dashed red line overlapping the mean profiles corresponds to the annual mean of the Mauna Loa measurements as in Figure 5, right, which confirms the accurate fit provided by kinetic model (15) only driven by fossil fuel emissions.

To be complete, we can roughly estimate the missing extrapolated contribution of the land-use change emissions $u_3$, the cyan curve in Figure 2, whose long-term average holds $\bar{u}_3 \simeq 0.55$ ppm/y. If the latter value is extrapolated by assuming future constant emissions (a conservative assumption since $u_3$ is decreasing), the steady state response of Eq. (15) holds

$$\bar{u}_3 \hat{\tau} \simeq 29 \text{ ppm} \sim 0.1 \hat{\underline{x}} \quad , \tag{35}$$

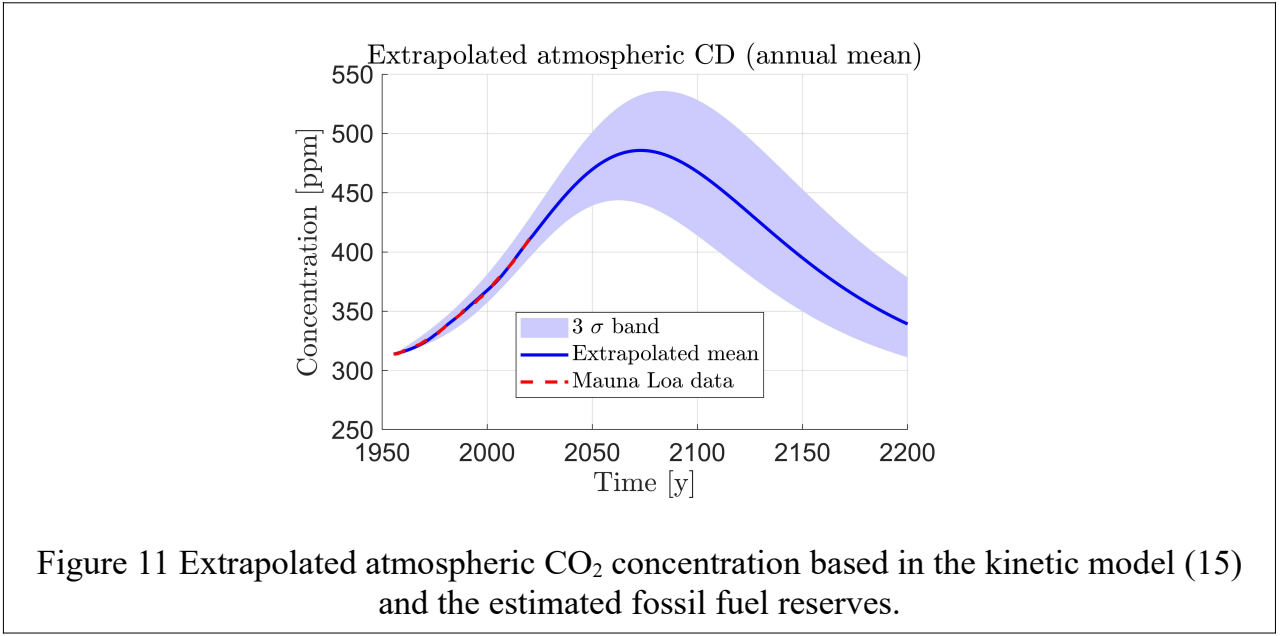

Figure 11 Extrapolated atmospheric $CO_2$ concentration based in the kinetic model (15) and the estimated fossil fuel reserves.

close to 10% of the estimated concentration equilibrium in Table 1.

We stress that no mitigation policy - like those aimed to keep pledges and targets each nation has signed - has been hidden in the kinetic model (15) and in the profile of the fossil fuel emissions. Therefore the profile band in Figure 11 should be kept as an upper limit of the atmospheric CO2 concentration under deregulated emissions (business-as-usual scenario).

The ranges of the concentration peak $x_{max}$ and its date $t_{max}$ are summarized in Table 4.

Table 4  Peaks and dates of the extrapolated atmospheric $CO_2$ concentration.

| No | Parameter | Symbol | Unit | Range | Comment |
|---|---|---|---|---|---|
| 1 | Peak | $x_{max}$ | ppm | 444~536 | |
| 2 | Date | $t_{max}$ | y | 2062.5~2083.5 | |

## 4.2 Comparison with literature projections

Before approaching comparison, we can summarize the paper development and results in Figure 10, Figure 11 and Table 4 by suggesting that the finite-reserve projections should be kept as the upper bounds of any other projection.

Comparison with the literature looks not immediate, as projections in [5], [6], [29] and [32], concern the $CO_2$ equivalent of the whole atmospheric GHGs, which as already noticed, follow different intake and removal mechanisms. Projections restricted to $CO_2$ appear in [34] but the relevant data files seem unavailable (the relevant datasets in [35] have been explored). We follow two roads.

We start by comparing the CAT (Climate Action Tracker) projections of the equivalent $CO_2$ emissions by greenhouse gases, as in [7], whose data are available. A caveat is that our projections in Figure 9 are restricted to fossil fuels. For this reason, graphical comparison in Figure 12 has been improved by downshifting the CAT current policies projections (solid and dashed red lines) of the GHGs emissions to overlap the uncertainty band (shaded area) of our extrapolations. According to OWID data, the constant down shift amounts to about $\Delta u = 13$ GtCO2/y, that is to the sum of methane and nitrous oxide emissions converted to $CO_2$ equivalent mass. A confirmation is given by the historical emission data (red and blue irregular curves) that overlap less a small drift. Overlap confirms that the current policy scenarios (upper and lower) coincide with our finite reserve scenario and as such should play the role of the upper bound of any other projection. For completeness, Figure 12 shows also the projection of a climate mitigation policy, the '2030 targets'.

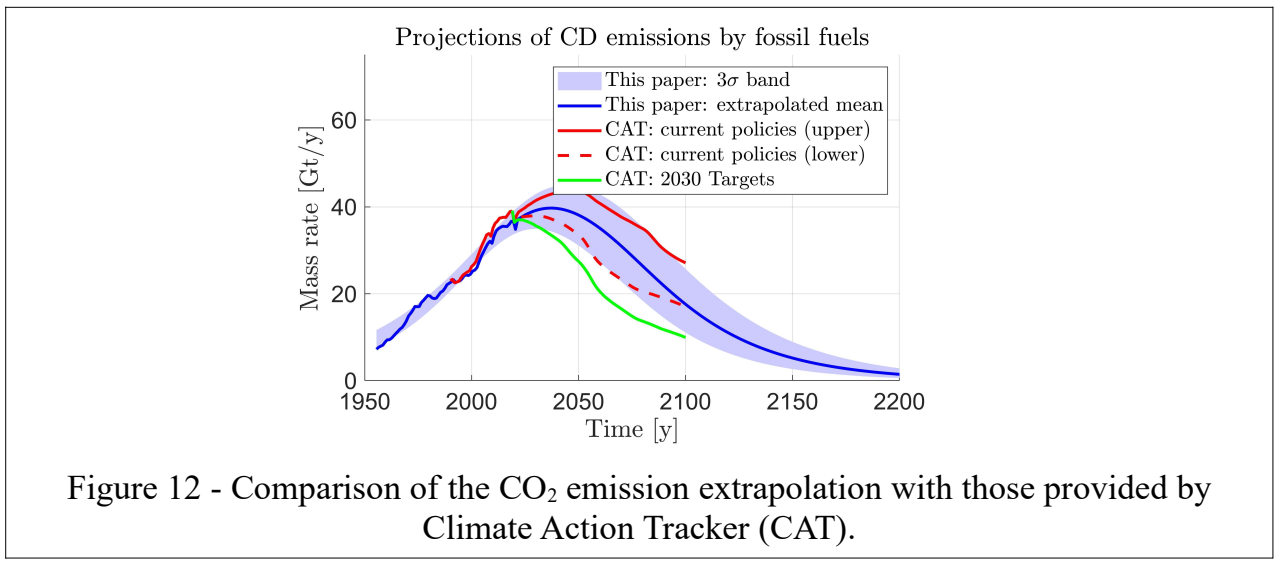

Figure 12 - Comparison of the $CO_2$ emission extrapolation with those provided by Climate Action Tracker (CAT).

The second set of comparisons is with IPCC projections in [34]. They should supersede previous IPCC and literature projections [40]. Their plots, restricted to $CO_2$ emissions and derived from [36], have been overlapped on Figure 10 and Figure 11.

Figure 13 compares the projection of the CO2 emission by fossil fuels in Figure 10 with IPCC projections (6$^{th}$ Assessment Report) of the five different scenarios. Historical data, though restricted to $CO_2$ emissions, differ by an offset of about 5 GtCO2/y. Profiles have been untouched unlike in Figure 12. The offset reason amounts to the land-use change emissions which are missing in the paper projections as already explained. Figure 5.5 in Chapter 5 of [34] shows, since 1960, a mean emission range 4.4~5.8 GtCO2/y , but with a profile rather different from that in Figure 2.

The five scenarios are explained in [34] (see also [37]). SSPx-y.y stands for Shared Socioeconomic Pathway, x=1 to 5 denotes the class of scenarios and y.y denotes the net radiative forcing [W/m$^2$] at year 2100. Radiative forcing is the name given by IPCC to the algebraic sum of natural (sun radiation change) and anthropogenic (GHG concentration change) exogenous radiant energy fluxes [W/m$^2$] which perturb the energy equilibrium of the Earth's biosphere and consequently the climate.

Needless to say, SSP 3-7.0 projection (high GHG emissions, $CO_2$ emissions double by 2100) and SSP 5-8.5 projection (very high GHG emissions, $CO_2$ emissions triple by 2075) look fully outside of the envelope defined by finite-reserve projections. SSP 2-4.5 projection (intermediate GHG emissions: $CO_2$ emissions around current levels until 2050, then falling but not reaching net zero by 2100) looks close to the mid profile of the CAT current policy projections in Figure 12, and therefore to the paper mean finite-reserve extrapolation.

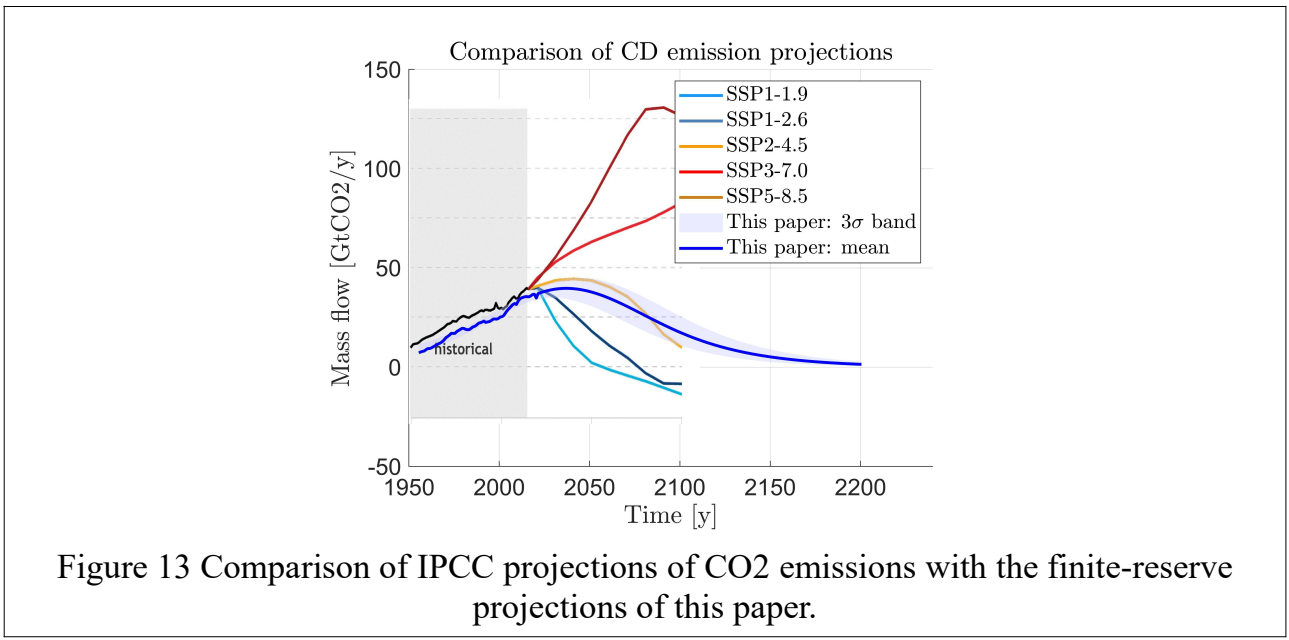

Figure 13 Comparison of IPCC projections of CO2 emissions with the finite-reserve projections of this paper.

Figure 14 shows the comparison of the IPCC projections of the atmospheric $CO_2$ concentration due to CO2 emissions in Figure 13 with the finite-reserve projections in Figure 11. At first sight, comparison looks coherent with Figure 13, in the sense that 1) SSP 3-7.0 and SSP 5-8.5 projections are fully outside of the envelope defined by the finite-reserve projections and 2) SSP 2-4.5 projection (orange color) drifts from the mean profile of this paper (blue color) because of the offset in Figure 13.

Actually, the drift deserves a deeper insight on the base of the key equation (15) and of their estimated parameters in Table 1. The offset value (between yellow and blue profiles in Figure 13) roughly amounts to $\Delta u \simeq \mu_{CO_2} 5$ GtCO2/y $= 0.64$ ppm/y (see (5)) and lasts for about $\Delta t = 100$ y before becoming negative. The resulting concentration perturbation $\Delta x(t)$ should stay well below the steady state $\Delta x_\infty = \hat{\tau} \Delta u \simeq 26 \sim 42$ ppm (see Table 1). Actually, the overshoot at $t = 2100$ happens to be larger than 100 ppm, raising questions about projection soundness. How their complex simulation packages have been validated? Complex dynamic models like Earth System and General Circulation Models employed by IPCC projections (see Annex II of [34]) are more prone to drift than simpler models. As a guess, an excess drift of this kind may be attributed to much longer time constants (hence to a larger steady state) than that estimated in Section 2. Questions may be raised about the short time interval (1955 to 2021) of the measurements compared to the estimated time constant $\hat{\tau}$ close to half century. Firstly, time duration is coded in the estimated uncertainty; secondly, extension to longer past intervals as in Figure 6, though affected by irregular measurements, has shown rather invariant estimates.

As a conclusion, we point out that simple yet validated by experimental data dynamic models, may pave the way for a better understanding, check and regulation of complex model projections.

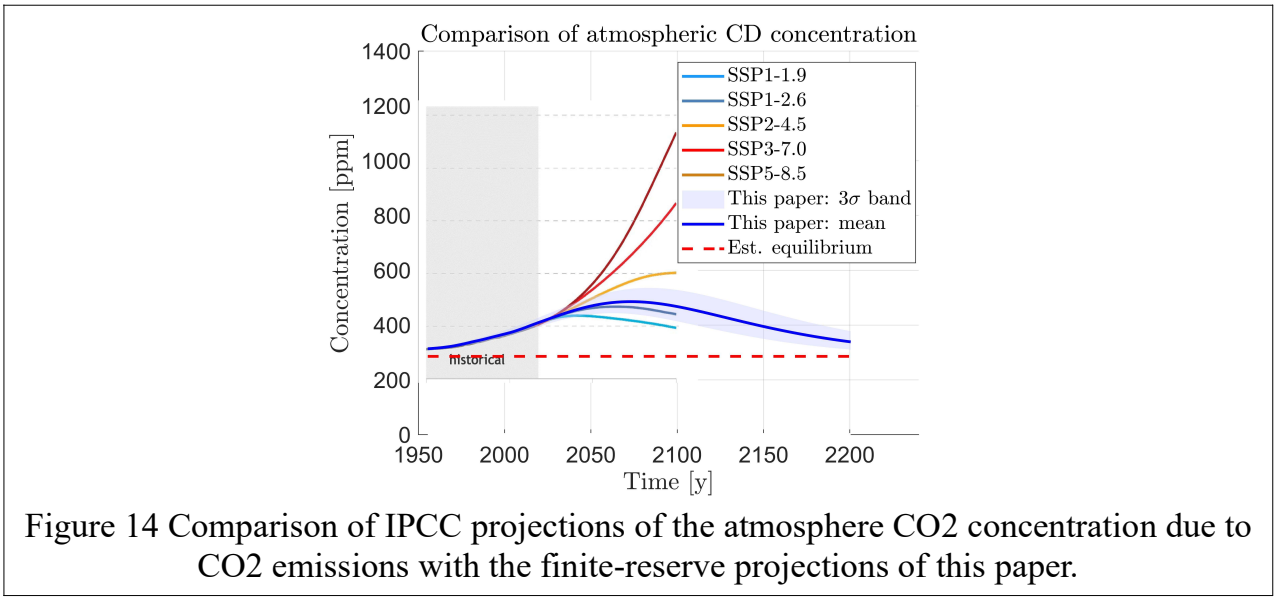

Figure 14 Comparison of IPCC projections of the atmosphere CO2 concentration due to CO2 emissions with the finite-reserve projections of this paper.

## 5  Conclusions

The paper starts from two observations: 1) the atmospheric $CO_2$ growth rate is smaller than that ascribed to the emission of fossil fuels combustion, 2) the fossil fuel reserves are finite. The first observation leads to a simple kinetic equation capable of accounting for the atmospheric $CO_2$ absorption by land and ocean. The second observation leads to a simple model for extrapolating the emission of single fossil fuels under current reserve constraint.

Driving the kinetic model by the extrapolated emissions allowed us to project the atmospheric CO2 concentration close to the zero-reserve epoch. The resulting mean profile and the relevant statistical bounds suggest an upper physical limit to complex simulations.

The method advantage is to stem from simple physical models, whose parameters can be easily estimated together with their uncertainty from available data. Integration in time and estimation procedures are so simple that students and scholars can easily and profitably repeat, check and update them.

The projections of fossil fuel emissions allows us to explicitly constrain their emission extrapolation to available reserves. Extension to GHG emission from non fossil sources was not the aim of the paper. Notwithstanding this limitation, a comparison with well-known projections of the atmospheric $CO_2$ concentration has favored, with a reasonable degree of confidence, a reciprocal check. As a result, IPCC projections look in many instances incompatible with the finite-reserve constrain. Therefore such projections should be considered as flawed, being inconsistent with the physical limits of fossil fuel exploitation

June 23, 2022, https://www.climate.gov/news-features/understanding-climate/climate-change-atmospheric-carbon-dioxide.

[37] B. O'Neill et al., The Scenario Model Intercomparison Project (ScenarioMIP) for CMIP6, *Geoscientific Model Development*, Vol. 9, 2016, pp. 3461–3482.

[38] V. Eyring et al., Overview of the couple model intercomparison project pahse 6 (CMIP6) experimental design and organization, *Geoscientific Model Development*, Vol. 9, No.5, 2016, pp. 1937-1958.

[39] R. Revelle and H. E. Suess, Crabon dioxide exchange between atmosphere and ocean and the question of an increase of atmospheric $CO_2$ during the past decades, *Tellus*, Vol. 9, No. 1, 1957, pp. 18-27.

[40] M. Meinshausen et al., The RCP greenhouse gas concentrations and their extensions from 1765 to 2300, *Climate Change*, Vol. 109, 2011, pp. 213-241.

[41] Intergovernmental panel on climate change (IPCC), *Working Group I: The Scientific Basis*, retrieved from https://archive.ipcc.ch/ipccreports/tar/wg1/016.htm, September 2022.

# 7   Appendix A – Derivation of the $CO_2$ reservoir state equations

## 7.1   Notations and assumptions

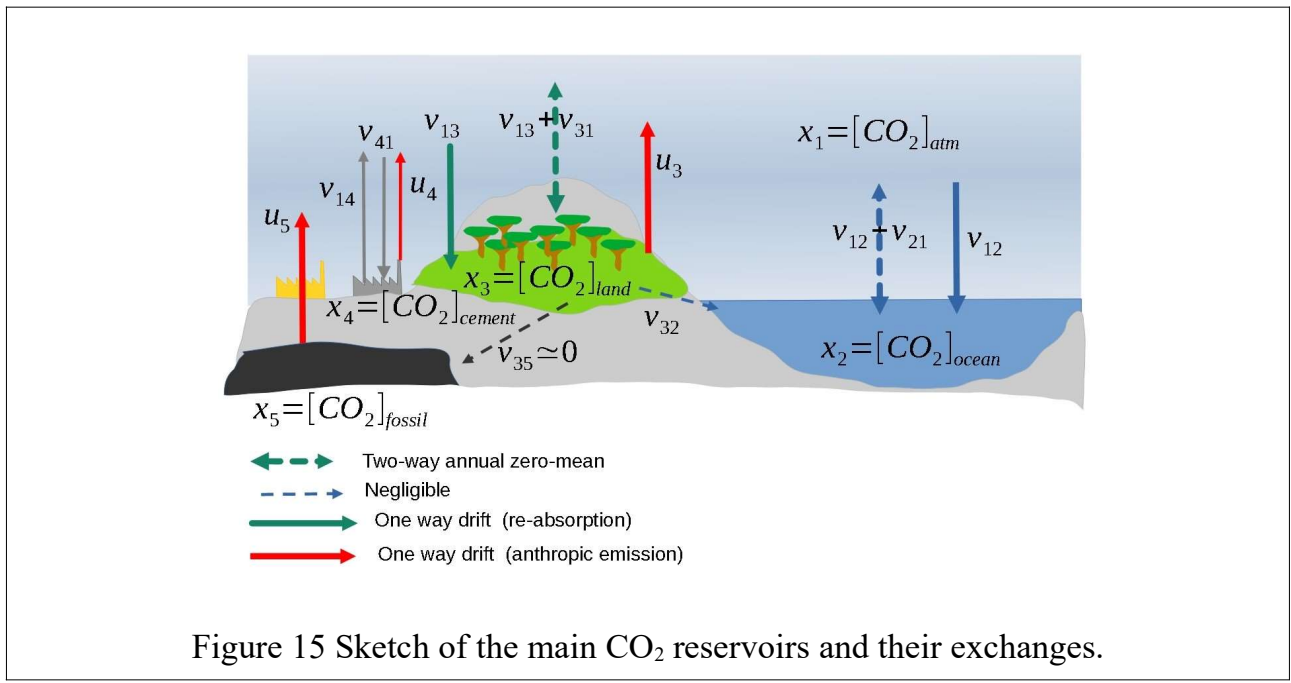

Figure 15 Sketch of the main $CO_2$ reservoirs and their exchanges.

Figure 15 sketches the different $CO_2$ reservoirs and their flows. The atmosphere is treated as a reservoir as well as the fossil fuel deposits. A further reservoir, to be neglected in our treatment, is the Earth's interior as it may emit $CO_2$ through volcanic eruptions. Artificial reservoirs capable of uptaking $CO_2$, though envisaged, are neglected.

Reservoirs are denoted by the subscript $s=1,...,5$, where $s=1$ refers to atmosphere, $s=2$ to ocean, $s=3$ to land, $s=4$ to cement and $s=5$ to the overall fossil fuel, including coal, oil and natural gas.

Given a time instant $t$, the amount of $CO_2$ existing in the reservoir $s$ is denoted by $x_s(t)$ and the time rate (uptaking if positive or emission if negative) by $\dot{x}_s(t)$. The sum $x_0(t)=\sum_{s=1}^{5} x_s(t)$ corresponds to the amount of $CO_2$ existing in the Earth's biosphere at time $t$, which, having neglected volcanic eruptions, must be kept as constant in time. In other words, we assume (Assumption 1) that

$$\text{Assumption 1: } \dot{x}_0(t)=0, \text{ at any } t \geq t_0 \ . \tag{36}$$

The assumption entrains that time variation of the reservoir state $x_s(t), h=1,...,5$ is only driven by the $CO_2$ exchange between them.

We distinguish between two kinds of exchange rates, natural and anthropogenic. Natural emission from reservoir $s$ to $h$, where $CO_2$ is uptaken, is denoted by $\pm v_{sh}(x_s(t))$. Negative sign refers to emission, positive to uptake. The series expansion of the exchange function around a constant equilibrium $\underline{x}_s$ can be truncated to the first-order as follows

$$v_{sh}(x_s(t)) \simeq v_{sh}(\underline{x}_s(t))+k_{sh}(x_s(t)-\underline{x}_s) \tag{37}$$

where $k_{sh}$ is the kinetic constant [1/s] of the exchange around the equilibrium $\underline{x}_s$. The first-order term has the same form of the direct and reverse reaction rates in (10) and (11), if we interpret the concentration $[CO_2]$ as a perturbation from equilibrium.

The annual zero-mean two-way component of the natural exchanges (say the carbon cycle of the biomass) is neglected by treating $x_s(t)$ as the current annual average of instantaneous reservoir levels as defined in (2).

The anthropogenic exchange is assumed (Assumption 2) to only occur between a reservoir $s \neq 1$ and the atmosphere $s=1$. The algebraic sum of emissions from and uptakes by $s \neq 1$ is denoted by $-u_s(t)$, which is negative when emissions dominate.

Anthropogenic exchanges (actually emissions) became significant since the dawn of the industrial era. Anthropogenic exchanges may also account for perturbations of the natural exchanges due to reduction/enlargement of the reservoir capacity. For instance, forestry reduction may diminish land capacity, and consequently its own $CO_2$ uptake rate. The relevant algebraic sum is collected under the name of land-use change emissions as shown in Figure 15.

## 7.2 Differential equations and equilibrium

The simplest equation is that of fossil fuels. The rate $\dot{x}_5$ is assumed (Assumption 3) to be only explained by anthropogenic emissions, which fact suggests the first-order differential equation

$$\text{Assumption 3: } \dot{x}_5(t)=-u_5(t), \ x_5(t_0)=x_{50}, \ u_5(t) \geq 0 \ , \tag{38}$$

where $x_{50}$ denotes the fossil fuel reserve at time $t_0$. Exchanges with reservoirs other than atmosphere are neglected.

The first-order differential equation of a generic reservoir $s=2,3,4$ explains the rate $\dot{x}_s(t)$ as the combination of the exchange with other reservoirs and of the anthropogenic emission $-u_s(t)$ to the atmosphere. We write

$$\dot{x}_s(t)=\sum_{h\neq s}\left(-v_{sh}(x_s)+v_{hs}(x_h)\right)-u_s(t),\ x_s(t_0)=x_{s0}\ . \tag{39}$$

The atmospheric equation is similar to (39), but includes all the anthropogenic emissions as follows

$$\dot{x}_1(t)=\sum_{h=2}^{4}\left(-v_{1h}(x_s)+v_{h1}(x_h)\right)+\sum_{h=2}^{5}u_h(t),\ x_1(t_0)=x_{10}\ . \tag{40}$$

The set of the equations (38), (39) and (40) may be referred to the as the carbon exchange equations of the biosphere. We can prove that $\dot{x}_0(t)=\sum_{h=1}^{5}\dot{x}_h(t)=0$, in agreement with the conservation equation (36). Carbon exchange equations can be simplified by neglecting the reciprocal exchange between $s=2,3,4$ (Assumption 4), namely

$$\text{Assumption 4: } v_{sh}(t)=0, s,h=2,3,4, s\neq h\ . \tag{41}$$

The simplified set of equations becomes

$$\begin{aligned}
\dot{x}_1(t)&=\sum_{h=2}^{4}\left(-v_{1h}(x_1)+v_{h1}(x_h)\right)+\sum_{h=2}^{5}u_h(t),\ x_1(t_0)=x_{10}\\
\dot{x}_2(t)&=-v_{21}(x_2)+v_{12}(x_1)-u_2(t),\ x_2(t_0)=x_{20}\\
\dot{x}_3(t)&=-v_{31}(x_3)+v_{13}(x_1)-u_3(t),\ x_3(t_0)=x_{30}\\
\dot{x}_4(t)&=-v_{41}(x_4)+v_{14}(x_1)-u_4(t),\ x_4(t_0)=x_{40}\\
\dot{x}_5(t)&=-u_5(t),\ x_5(t_0)=x_{50}
\end{aligned} \tag{42}$$

The equilibrium state is obtained by zeroing, in (42), the anthropogenic exchanges $u_h(t)=0$, and the rate $\dot{x}_h(t)=0$. A pair of equilibriums can be arbitrarily chosen, for instance the fossil fuel $\underline{x}_5$ and the overall amount $\underline{x}_0=\sum_{h=1}^{5}\underline{x}_h$. This looks reasonable since they cannot be fixed by reciprocal exchange, being decided by past history. By replacing the overall amount $\underline{x}_0$ with the atmospheric equilibrium $\underline{x}_1$, ocean, land and cement levels can be found from the exchange equilibriums:

$$v_{s1}(\underline{x}_s)=v_{1s}(\underline{x}_1),\ s=2,3,4 \tag{43}$$

## 7.3 Perturbation equation around equilibrium

Let us denote the column vector of the reservoir states with $x=[x_1,x_2,x_3,x_4,x_5]$, where the inline notation has been used. The equilibrium vector is denoted with $\underline{x}$ and the perturbation from the equilibrium with $\delta x = x - \underline{x}$. Clearly $\delta \dot{x} = \dot{x}$. The anthropogenic exchanges $u_s$ are collected into the column vector $u=[u_2,u_3,u_4,u_5]$. Replacement of the expansion (37) into (42) and cancellation of the constant part in (37) (it vanishes because of the equilibrium), provides a system of first-order differential equations, written in the following matrix form:

$$\delta \dot{x}(t) = A \delta x(t) + B u(t), \quad \delta x(t_0) = \delta x_0, \tag{44}$$

where equation matrices hold

$$A = \begin{bmatrix} -(k_{12}+k_{13}+k_{14}) & k_{21} & k_{31} & k_{41} & 0 \\ k_{12} & -k_{21} & 0 & 0 & 0 \\ k_{13} & 0 & -k_{31} & 0 & 0 \\ k_{14} & 0 & 0 & -k_{41} & 0 \\ 0 & 0 & 0 & 0 & 0 \end{bmatrix}, \quad B = \begin{bmatrix} -1 & 0 & 0 & 0 \\ 0 & -1 & 0 & 0 \\ 0 & 0 & -1 & 0 \\ 0 & 0 & 0 & -1 \end{bmatrix}. \tag{45}$$

The parameters $k_{1s}, s=2,3,4$ in the first row of $A$ correspond to the direct kinetic constant $k_{dir}$ in (10). $k_{12}$ refers to the reaction between atmospheric $CO_2$ and ocean carbonic acid as in (8), $k_{s3}$ refers to the photosynthesis reactions between atmospheric $CO_2$ and vegetation and finally $k_{s4}$ to cement carbonation. They in turn become the reverse kinetic constants of the inverse reactions in the rows $s>1$. The parameters $k_{s1}, s=2,3,4$ in the first row correspond to $k_{inv}$ in (11), namely to the reverse kinetic constants of the reactions from atmospheric $CO_2$ to ocean, vegetation and cement compounds where carbon is confined. They in turn become the direct kinetic constants of the relevant inverse reactions in the rows $s>1$.

Equation (13) follows from (44) by assuming (Assumption 5) that during the equation integration interval $t_0 \le t < t_{End}$ the carbon level of ocean, land and cement remains constant and equal to equilibrium, that is

$$\text{Assumption 5:} \quad x_s(t) = \underline{x}_s, \quad s=2,3,4 \Rightarrow \delta x_s(t) = 0. \tag{46}$$

Under this assumption, equation (44) reduces to a pair of equations, that of the atmospheric reservoir in (15) and that of the fossil fuel in (27). From (46), by writing $u=u_2+u_3+u_4+u_5$, by denoting the direct kinetic constant as $k=k_{12}+k_{13}+k_{14}$, by making explicit the equilibrium $\underline{x} = (k_{21}\underline{x}_2 + k_{31}\underline{x}_3 + k_{41}\underline{x}_4)/k$, by dropping the subscript $s=1$ (atmosphere) and by recalling the fossil fuel notations $r=x_5, c=u_5$ of Section 3.1, we obtain

$$\begin{aligned}\dot{x}(t) &= -k(x(t)-\underline{x})+u(t), x(t_0)=x_0 \\ \dot{r}(t) &= -c(t), r(t_P)=r_P, t_0<t_P<t_{End}\end{aligned} \quad (47)$$

The fossil fuel initial time $t_P$ corresponds to the present epoch, denoted with $t_N$ in Section 3.1, which implies that the initial state $r_P$ represents the present fossil reserves, $x_0$ is the atmospheric concentration at $t=t_0$ and the linear feedback term $-k(x(t)-\underline{x})$ under $x(t)>\underline{x}$ is the *ocean/land absorption flow*.
Deviation from Assumption 5 in (46) is such to affect the first equation in (47) with a model error written as the sum of a bias $\Delta \underline{v}_{234}=k\Delta \underline{x}_{234}$ and a zero-mean variable term $w_{234}(t)$. The bias is absorbed by the unknown equilibrium $\underline{x}$, whereas $w_{234}(t)$ is absorbed by the measurement error of the regression equation (18).

## 8  Appendix 2 – List of historical data, their source and uncertainty

Table 5 shows the source of the historical data employed in this paper. The standard deviations of rows 2, 3 and 4 come from the Global Carbon Budget (see [12] and the comments on the data file). The standard deviation in row 1 comes from the atmospheric $CO_2$ annual mean assigned by NOAA to Mauna Loa data [15] since 1959. The standard deviation of ice core data is reported to hold 0.1 ppm in [18].

| Table 5 - Historical data employed in the paper | | | | | |
|---|---|---|---|---|---|
| No | Variable (equation) | Symbol | Unit | Standard deviation | Source and file |
| 1 | Atm $CO_2$ annual mean (20) | $\check{x}$ | $\frac{\mu \text{mol}}{\text{mol}}=\text{ppm}$ | 0.12 (NOAA), 0.1 (ice cores) | SCRIPPS Program, merged_ice_core_yearly.csv [17] (see also [18] for ice core data) |
| 2 | Atm $CO_2$ annual rate (18) | $\Delta\check{x}$ | $\frac{\text{GtCO2}}{\text{y}}\left(\frac{\text{ppm}}{\text{y}}\right)$ | 0.20 (0.095) | ICOS, Data supplement to the GBC 2021 [12] and [16], Global_Carbon_Budget_2021v1.0.xlsx, sheet Historical Budget |
| 3 | Land-use change emissions (35) | $\check{u}_3$ | $\frac{\text{GtCO2}}{\text{y}}\left(\frac{\text{ppm}}{\text{y}}\right)$ | 0.7 (0.33) | Same. |
| 3 | Fossil fuel emissions (18), (20) | $\check{u}=\check{c}$ | $\frac{\text{GtCO2}}{\text{y}}$ | 5% (0.1 to 0.5) | Same |

| 4 | Fuel emission by type (33) | $\check{c}_f$ | $\frac{GtCO2}{y}$ | NA | OWID, co2-emissions-by-fuel-line (World lines)4 |
| 5 | Coal reserves (30) | $r_c$ | 1000 tonnes | NA | OWID, coal_proved_reserves.csv (World lines) |
| 6 | Oil reserves (30) | $r_o$ | barrels | NA | OWID, oil_proved_reserves.csv (World lines) |
| 7 | Natural gas reserves (30) | $r_g$ | m$^3$ | NA | natural_gas_proved_reserves.csv, OWID (World lines) |